\documentclass[12pt]{article}%
\usepackage{amsmath}
\usepackage{amsfonts}
\usepackage{amssymb}
\usepackage{graphicx}
\usepackage{hyperref}%
\ifx\pdfoutput\relax\let\pdfoutput=\undefined\fi
\newcount\msipdfoutput
\ifx\pdfoutput\undefined\else
\ifcase\pdfoutput\else
\ifx\paperwidth\undefined\else
\ifdim\paperheight=0pt\relax\else\pdfpageheight\paperheight\fi
\ifdim\paperwidth=0pt\relax\else\pdfpagewidth\paperwidth\fi
\fi\fi\fi
\begin{document}

\title{Renormalization of a modified gravity with a quadratic Riemann tensor term}
\author{M. Chaves\\\textit{Universidad de Costa Rica}\\\textit{San Jos\'{e}, Costa Rica}\\mchaves@cariari.ucr.ac.cr}
\date{July 25, 2008}
\maketitle

\begin{abstract}
We consider a modified form of gravity, which has an extra term quadratic in
the Riemann tensor. This term mimics a Yang-Mills theory. The other defining
characteristic of this gravity is having the affine connection independent of
the metric. (The metricity of the metric is rejected, too, since it implies a
Levi-Civita connection.) It is then shown that, in the low density limit, this
modified gravity does not differ from the General Theory of Relativity. We
then point out that its Lagrangian does not contain partials of the metric, so
that the metric is not a quantum field, nor does it contribute propagators to
the Feynman diagrams of the theory. We also point out that the couplings of
this theory (that determine the topological structure of the Feynman diagrams)
all come from the term quadratic in the Riemann tensor. As a result of this
situation, the diagrams of this theory and the diagrams of a Yang-Mills theory
all have the same topology and degree of divergence, up to numerical
coefficients. Since Yang-Mills theories are renormalizable, it follows that
this theory should also be renormalizable.

\end{abstract}



\section{INTRODUCTION.}

There has been much interest in the last few years on the subject of
modifications of the action of the General Theory of Relativity (GTR)%
\begin{equation}
S_{\text{GR}}=\int\sqrt{-g}\left(  \frac{2}{\kappa^{2}}R+\mathcal{L}%
_{M}\right)  \,d^{4}x, \label{lag1}%
\end{equation}
where $\kappa=\sqrt{32\pi G}$ is the gravitational coupling constant and
$\mathcal{L}_{M}$ is the matter and radiation Lagrangian.\cite{SF} The main
motivation for this interest comes from the questions posed by astronomical
observations that suggest the existence of dark matter and dark
energy.\cite{NO} Here we are going to consider a modification to the
Lagrangian of the GTR, but our interest lies in improving its renormalization
prospects. We use a particular combination of old ideas that result, taken
together, in a new approach to this subject.

There are two fundamental difficulties with attempts at renormalizing the GTR.
The first difficulty is that the gravitational coupling constant $\kappa
\sim1/E_{\text{Planck}}$ has units of reciprocal energy $E^{-1}$. This should
lead to a quantum field theory that cannot be renormalized since, as the
number of loops increases, the order of the divergence must increase, too, and
then there has to be an infinite number of distinct renormalization
counterterms. The second difficulty has to do with the fact that, since the
Levi-Civita connection is expressed as a sum of partials of the metric, then
there are going to be diagrams in the theory that contain powers of first- and
second-order partials of the metric.\cite{Stelle} This situation leads to
double quadratic poles in the propagators and the combinations of these
propagators can involve negative probabilities which make the theory
meaningless.\cite{KN}

One way of addressing the first difficulty is to use a theory for gravity that
does not use a coupling constant with units. Since the Riemann tensor has
units of $E^{-2}$ a term in the Lagrangian involving some squared form of the
Riemann tensor would have units of $E^{-4}$ and thus would require a coupling
constant with no units.\cite{F2} Different possibilities for a squared Riemann
tensor Lagrangian\cite{SS} are of the form $R^{2},$ $R_{\mu\nu}R^{\mu\nu},$
and $R_{\mu\nu\xi\rho}R^{\mu\nu\xi\rho}$ (for purposes of inclusion in an
action these terms are not independent since they are related by the
Gauss-Bonnet topological invariant). Unfortunately, having a coupling constant
with no units does not solve all the quantization problems. Powers of the
first- and second-order partials of the metric still appear, and new types of
diagrams keep showing up at higher orders, that require the inclusion of an
infinite number of counterterms not reducible to a finite set of primitive diagrams.

String theory does contain, as an effective quantum field theory, an infinite
number of terms of these type. In principle, string theory can be replaced by
an infinite number of local fields. By integrating out of the path integral
the massive fields it should be possible to achieve an effective non-local
field theory. Not surprisingly in practice only approximate methods can be
used for this purpose. One such method is to construct the quantized string
theory on background fields, and then use conformal invariance to build up
constraints between the background fields, which are then identified with the
$\beta$ functions of the corresponding $\sigma$ model.\cite{strings1}
Alternatively, it is possible from string theory to calculate the scattering
amplitudes of the massless particles in tree approximation, and from there
guess what the effective Lagrangian for the massless particles can be. The
higher-order terms in the Lagrangian can be constructed by iteration from the
first Lagrangian, using higher-order string interactions and unitarity
considerations.\cite{strings2} The resulting effective theory includes
metrics, antisymmetric tensors, dilaton fields and fermion fields, as well as
gauge background fields in the case of the heterotic string.

In this low-energy interpretation given to the string theory, the terms in the
metric and partials of the metric are grouped to form Riemann tensors
constructed from Levi-Civita connections. The problems involved in the
quantization of an effective theory of this type are basically the same as
those encountered in a direct quantization of Einstein's gravity. This
situation obviously it does not constitute evidence of any fundamental
incorrectness of string theory, since we are dealing only with an
approximation to the much better behaved worldsheet, but, be it as may, we
cannot quantize directly this string-suggested theory of gravity, either.

In order to avoid the problems presented by the presence of powers of the
metric and its partials we resort to a metric-affine setup, also sometimes
also called Einstein-Palatini. We take the metric and the connection to be
completely independent. (The connection is taken to be symmetric and there is
no torsion.) Thus the connection is not Levi-Civita. The metric-affine setup
is often used as a prelude to establishing a gauge theory of
gravity,\cite{Phys Rep} but we will not proceed this way here. We are assuming
invariance only under coordinate diffeomorphisms.

As we shall see, the connection will still turn out to be Levi-Civita, but
only in the low-density limit, and as a result of the equations of motion. By
low-density it is meant that the energy and momentum densities are roughly of
the order of the average density of our universe or less.

Consider the modified action%
\begin{equation}
S_{\text{mod}}=\int\sqrt{-g}\left(  \frac{1}{4}R^{\rho}{}_{\sigma\mu\nu}%
g^{\mu\tau}g^{\nu\upsilon}R^{\sigma}{}_{\rho\tau\upsilon}+\frac{2}{\kappa^{2}%
}R+\mathcal{L}_{\text{M}}\right)  \,d^{4}x, \label{lag2}%
\end{equation}
which has an extra term with respect to (\ref{lag1}). The new term mimics the
curvature term in a Yang-Mills Lagrangian, with the group indices of the gauge
field replaced by coordinate indices; this reason for this choice over the
other quadratic forms will eventually become clear. This action generates two
equation of motion, one due to a first variation with respect to the
connection, and another due to a first variation with respect to the metric.

In the remainder of this paper we shall show that, while for low densities
this action cannot be distinguished from the GTR, it does have good
perspectives of renormalization.

\section{The metricity of the metric.}

The Levi-Civita connection depends on the metric:%
\begin{equation}
\Gamma^{\lambda}{}_{\mu\nu}=\frac{1}{2}g^{\lambda\xi}(g_{\mu\xi,\nu}+g_{\xi
\nu,\mu}-g_{\mu\nu,\xi}). \label{LC}%
\end{equation}
This relation between the connection and the metric follows from the so-called
metricity of the metric,%
\begin{equation}
g_{\mu\nu;\lambda}=0. \label{metricity}%
\end{equation}
Since we are not taking the connection to be Levi-Civita, the metric has no
metricity, that is, (\ref{metricity}) does not hold. The metricity of the
metric clearly holds in the 3-dimensional world of our experience, and,
theoretically, Einstein-Cartan theories usually respect metricity. So let us
study for a moment the implications of non-metricity.

The point of view on metricity in Riemannian geometry becomes clear in the
following representative definition, standard in textbooks on the subject. The
metric is an inner symmetric product $\left\langle \;,\,\right\rangle $
defined on the tangent space $TM_{p}^{(n)}$ that is associated with each point
$p$ of a real differentiable manifold of dimension $n$. A Riemannian manifold
is a manifold with a metric. In terms of a coordinate basis this product has
components%
\[
g_{\mu\nu}=\left\langle \frac{\partial}{\partial x^{\mu}},\frac{\partial
}{\partial x^{\nu}}\right\rangle =\left\langle \partial_{\mu},\partial_{\nu
}\right\rangle .
\]
Thus two vectors $\mathbf{u}$ and $\mathbf{v}$ have the metric or inner
product%
\[
\left\langle \mathbf{u,v}\right\rangle =\left\langle u^{\mu}\partial_{\mu
},v^{\nu}\partial_{\nu}\right\rangle =u^{\mu}g_{\mu\nu}v^{\nu}.
\]
Let us assume a covariant differentiation operator $\mathbf{\nabla.}$ Then the
standard definition for the covariant differentiation of the inner product of
two vector fields $\mathbf{u}$ and $\mathbf{v,}$ both defined along a
parametrized curve $x(t),$ is given by:%
\begin{equation}
\frac{d}{dt}\left\langle \mathbf{u,v}\right\rangle =\left\langle
\frac{\mathbf{\nabla u}}{dt}\mathbf{,v}\right\rangle +\left\langle
\mathbf{u,}\frac{\mathbf{\nabla v}}{dt}\right\rangle . \label{d-metric}%
\end{equation}
Using coordinates this definition immediately implies (\ref{metricity}). To
see this let us write this equation in components, using a coordinate frame.
Then (\ref{d-metric}) would have the following equivalent form in the open
coordinate patch that contains path $x(t)$:%
\begin{equation}
\frac{\partial}{\partial x^{\lambda}}(u^{\mu}g_{\mu\nu}v^{\nu})=(u^{\mu}%
{}_{,\lambda}+\Gamma_{\lambda\xi}^{\mu}u^{\xi})g_{\mu\nu}v^{\nu}+u^{\mu}%
g_{\mu\nu}(v^{\nu}{}_{,\lambda}+\Gamma_{\lambda\xi}^{\nu}v^{\xi}).
\label{e-metric}%
\end{equation}
Applying the rule for partial differentiation of a product on the left-hand
side of this equation, which results in three terms, we see metricity follows immediately.

But one can argue differently. One can say that the original definition
(\ref{d-metric}) is actually missing a term on the right-hand side of the
equation, because it is ignoring the possibility that the metric can also be
affected by the differentiation process. This becomes evident using
components. In this case the definition should include a covariant
differentiation of the metric, too, and the differentiation of the scalar
$u^{\mu}g_{\mu\nu}v^{\nu}$ results in%
\[
\frac{\partial}{\partial x^{\lambda}}(u^{\mu}g_{\mu\nu}v^{\nu})=u^{\mu}%
{}_{;\lambda}g_{\mu\nu}v^{\nu}+u^{\mu}g_{\mu\nu;\lambda}v^{\nu}+u^{\mu}%
g_{\mu\nu}v^{\nu}{}_{;\lambda}.
\]
After we expand the covariant derivatives, all terms cancel and we are left
with the tautology $0=0,$ so that metricity is not implied. Thus definition
(\ref{d-metric}) is really equivalent to equation (\ref{metricity}).

In physics metricity is derived from Einstein's Equivalence Principle.
According to the Equivalence Principle, given a small region in spacetime, it
is always possible to find there a coordinate system that approximates a
locally inertial frame. Thus in the small region the metric is approximately a
Minkowski metric $\eta_{\mu\nu}=\operatorname{diag}(-1,+1,+1,+1),$ with null
first-order partial derivatives, but not necessary null second-order partial
derivatives. Therefore $\eta_{\mu\nu;\lambda}=0.$ We can transform this
equation to a different coordinate system, following the rules for
transforming tensors, and the result is then that, in all possible coordinate
systems, $g_{\mu\nu;\lambda}=0$. Thus if we do not want metricity we also have
to do away with the Equivalence Principle.\cite{Olmo} This principle gives
conceptual unity to the GTR and helps derive, up to a point, its equations,
but, since we are questioning the GTR, it is only to be expected that we
should also question its guiding principle. In this paper metricity turns out
to be true only as a low density limit.

\section{The low-density limit of the modified action.}

In this section we look at the low-density limit of Lagrangian (\ref{lag2}).
This limit is the state most of the Universe we can observe is in.

Let us compare, in this limit, the two terms which appear in (\ref{lag2}) that
involve the Riemann tensor. The first goes as $\bar{R}^{2},$ where $\bar{R}$
is the average value of the curvature scalar in our universe, and the second
as $\bar{R}L_{\text{Planck}}^{-2}.$ Since $|\bar{R}|\ll L_{\text{Planck}}%
^{-2},$ in this limit the inequality%
\[
\frac{1}{4}R^{\rho}{}_{\sigma\mu\nu}g^{\mu\tau}g^{\nu\upsilon}R^{\sigma}%
{}_{\rho\tau\upsilon}\ll\frac{2}{\kappa^{2}}|\bar{R}|
\]
holds, and Lagrangian (\ref{lag2}) can be approximated by the Lagrangian that
appears in (\ref{lag1}), that is, the GTR Lagrangian, except that we are still
considering the connection and the metric as independent. The first-order
variation of (\ref{lag1}) with respect to the connection results in Einstein's
equation, and the first-order variation with respect to the metric results in
metricity, as is well-known from the Einstein-Palatini
variation.\cite{Weinberg} Thus the resulting theory is very similar to the GTR.

\section{Improved renormalization prospects of the Lagrangian.}

In this section we study Lagrangian (\ref{lag2}), where the Riemann tensor is
a functional of the connection given by%
\begin{equation}
R^{\rho}{}_{\sigma\mu\nu}=\Gamma^{\rho}{}_{\nu\sigma,\mu}-\Gamma^{\rho}{}%
_{\mu\sigma,\nu}+\Gamma^{\rho}{}_{\mu\eta}\Gamma^{\eta}{}_{\nu\sigma}%
-\Gamma^{\rho}{}_{\nu\eta}\Gamma^{\eta}{}_{\mu\sigma}. \label{Riemann}%
\end{equation}
We are interested in showing that the theory associated with (\ref{lag2}) is renormalizable.

The two fields to quantize are the metric $g_{\mu\nu}$ and the connection
$\Gamma^{\rho}{}_{\mu\nu},$ considered independent of each other. A
fundamental point is to notice that Lagrangian (\ref{lag2}) does not contain
partials of the metric, and that therefore there is no canonical momentum
field conjugate to the metric in this theory$.$ \emph{Thus the metric is not a
quantum field, just a classical background field.} One natural way to proceed
is to integrate it out of the effective action path integral; probably it
should be possible to do it. Here we will follow for now another procedure.

The elements involved in the diagrams pertaining the renormalization process
of a theory are basically vertices and propagators. In the theory we are
studying the connections have propagators \textit{but the metrics do not,}
since they are not quantum excitations. There are no partials of the metric
available in the action to allow the construction of a propagator. This means
that metrics, in the Feynman diagrams of the theory, will have no lines
associated with them. The vertices of the theory do involve both connections
and metrics. However, the metrics in the vertices serve only a bookkeeping
purpose, to keep track of the covariance or contravariance of an index, but no
line is connected to them.

In the Feynman diagrams of a theory, some of the diagrams involve closed
loops, and some of these closed loops are divergent. If the more complicated
divergent diagrams can be built up out of primitive subdiagrams (where all the
primitive subdiagrams are members of a finite set), and the resulting diagrams
are then non-divergent (discounting the divergent primitive subdiagrams, for
which there are counterterms available in the Lagrangian) we then call the
theory renormalizable. We proceed to show that that is precisely the case for
this theory of gravity.

Let us write below the Lagrangian of action (\ref{lag2}) with the Riemann
tensor expressed in terms of the connection:%
\begin{align}
\mathcal{L}_{\text{mod}}  &  =\frac{1}{4}\left(  \Gamma^{\rho}{}_{[\nu
\sigma,\mu]}+\Gamma^{\rho}{}_{[\mu\eta}\Gamma^{\eta}{}_{\nu]\sigma}\right)
g^{\mu\tau}g^{\nu\upsilon}\left(  \Gamma^{\sigma}{}_{[\upsilon\rho,\tau
]}+\Gamma^{\sigma}{}_{[\tau\eta}\Gamma^{\eta}{}_{\upsilon]\rho}\right)
\nonumber\\
&  +\frac{2}{\kappa^{2}}\left(  g^{\sigma\nu}\Gamma^{\rho}{}_{[\nu\sigma
,\rho]}+g^{\sigma\nu}\Gamma^{\rho}{}_{[\rho\eta}\Gamma^{\eta}{}_{\sigma]\nu
}\right)  +\mathcal{L}_{\text{M}}. \label{lag4}%
\end{align}
Notice the similarity of the first term with a Yang-Mills Lagrangian. The
Yang-Mills field is a vector field $A_{\mu a},$ where $a$ takes value in the
adjoint representation of a Lie algebra, while $\Gamma^{\rho}{}_{\mu\sigma}$
is an affine connection invariant under diffeomorphisms.

Notice, too, that the second term of (\ref{lag4}) contains only one- and
two-point coupling vertices. Remember here that the only quantum field of this
theory is the connection, and that this field appears linearly and
quadratically in the second term of (\ref{lag4}). Let us ignore this second
term, and deal with it at the end of this section. If we keep in our theory of
gravity only the first term in (\ref{lag4}), plus the matter and radiation
term $\mathcal{L}_{\text{M}},$ then the couplings that determine the topology
of a Feynman diagram are very similar to the ones of a Yang-Mills theory.

We can make transparent the similarity between both Lagrangians (gravity and
Yang-Mills), by using a trick. It consists in writing the affine connection so
that it looks like a Yang-Mills field. Consider the connection $\Gamma^{\rho
}{}_{\sigma\nu}$ as four matrices $\Gamma_{\sigma},$ $\sigma=0,1,2,3,$ each
one with 16 components classified by two indices $\rho$ and $\nu,$ the first
contravariant and the second covariant:%
\[
\Gamma^{\rho}{}_{\sigma\nu}=\left(  \Gamma_{\sigma}\right)  ^{\rho}{}_{\nu}.
\]
Using this convention we can write, for instance,%
\[
\Gamma^{\rho}{}_{\mu\eta}\,\Gamma^{\eta}{}_{\nu\sigma}\,\delta^{\sigma}%
{}_{\rho}=(\Gamma{}_{\mu}\,\Gamma{}_{\nu})^{\rho}{}_{\sigma}\,\delta^{\sigma
}{}_{\rho}=\operatorname{Tr}(\Gamma{}_{\mu}\,\Gamma{}_{\nu}).
\]
Keeping these ideas in mind we can express the first term of (\ref{lag4}) as
follows:%
\begin{equation}
\mathcal{L}_{\text{1%
${{}^o}$
term}}=\frac{1}{4}\operatorname{Tr}\left(  (\Gamma{}_{[\nu,\mu]}+[\Gamma
{}_{\mu},\Gamma{}_{\nu}])g^{\mu\tau}g^{\nu\upsilon}(\Gamma_{\lbrack
\upsilon,\tau]}+[\Gamma_{\tau},\Gamma_{\upsilon}])\right)  , \label{high}%
\end{equation}
in complete analogy with the Yang-Mills case.

The kinetic energy of the connection is given by some of the terms of
(\ref{lag4}):%
\[
\mathcal{L}_{\text{KE}}=\frac{1}{4}\Gamma^{\rho}{}_{[\nu\sigma,\mu]}g^{\mu
\tau}g^{\nu\upsilon}\Gamma^{\sigma}{}_{[\upsilon\rho,\tau]}.
\]
There is another very interesting and useful similarity between both types of
theories. In Yang-Mills theories frequently the following term is added to the
Lagrangian:%
\[
\mathcal{L}_{\text{YM fix}}=-\frac{1}{2\xi}(\partial_{\mu}A_{a}{}^{\mu})^{2}.
\]
It is evident that this term is not invariant under a Yang-Mills gauge
transformation and thus serves to partially fix the gauge and make the
propagator matrix non-singular so it can be inverted. In the theory of gravity
discussed here a similar term can be introduced for the same purpose. It is%
\begin{equation}
\mathcal{L}_{\text{grav fix}}=-\frac{1}{2\xi}\operatorname{Tr}\left(  \left(
\partial^{\mu}\Gamma{}_{\mu}\right)  ^{2}\right)  =-\frac{1}{2\xi}%
(g^{\lambda\mu}\partial_{\lambda}\Gamma^{\rho}{}_{\mu\sigma})(g^{\tau\nu
}\partial_{\tau}\Gamma^{\sigma}{}_{\nu\rho}). \label{fix}%
\end{equation}
The remarkable point is that this term is not invariant under diffeomorphisms
since $\partial_{\lambda}\Gamma^{\rho}{}_{\mu\sigma}$ is not a tensor; that is
why it serves to fix the gauge.

We can use (\ref{high}) and (\ref{fix}) to write the action of the kinetic
energy of the connection:%
\begin{align*}
&  \int\left(  \mathcal{L}_{\text{KE}}+\mathcal{L}_{\text{grav fix}}\right)
\,d^{4}x\\
&  =\frac{1}{2}\int\Gamma^{\rho}{}_{\mu\sigma}\left[  g^{\mu\nu}\partial
^{2}-\left(  1-\frac{1}{\xi}\right)  \partial^{\mu}\partial^{\nu}\right]
\Gamma^{\sigma}{}_{\nu\rho}\,d^{4}x\\
&  =-\frac{1}{2}\int\Gamma^{\rho}{}_{\mu\sigma}(x)D(x,y)^{\sigma\mu}{}_{\rho
}|^{\varsigma\nu}{}_{\tau}\Gamma^{\tau}{}_{\nu\varsigma}(y)\,d^{4}x{}d^{4}y,
\end{align*}
where the matrix $D(x,y)^{\sigma\mu}{}_{\rho}|^{\varsigma\nu}{}_{\tau}$ is
defined by%
\begin{align*}
D(x,y)^{\sigma\mu}{}_{\rho}|^{\varsigma\nu}{}_{\tau}  &  \equiv\delta^{\sigma
}{}_{\tau}\,\delta^{\varsigma}{}_{\rho}\left(  g^{\mu\nu}\partial^{2}-\left(
1-\frac{1}{\xi}\right)  \partial^{\mu}\partial^{\nu}\right)  \delta^{4}(x-y)\\
&  =\frac{1}{(2\pi)^{4}}\int\left[  \delta^{\sigma}{}_{\tau}\,\delta
^{\varsigma}{}_{\rho}\left(  g^{\mu\nu}(p^{2}-i\epsilon)-\left(  1-\frac
{1}{\xi}\right)  p^{\mu}p^{\nu}\right)  \right]  e^{ip\cdot(x-y)}{}d^{4}p.
\end{align*}
The propagator is the reciprocal of the matrix in the square brackets in the
equation above. This reciprocal actually exists since the matrix in the square
brackets is not singular due to the addition of the diffeomorphism-breaking
term (\ref{fix}).

Let us compare this result with the one for a Yang-Mills theory. The terms in
the Yang-Mills action for the propagator are%
\begin{align}
&  \frac{1}{2}\int A_{\mu a}\left[  g^{\mu\nu}\partial^{2}-\left(  1-\frac
{1}{\xi}\right)  \partial^{\mu}\partial^{\nu}\right]  A_{\nu a}\,d^{4}x\\
&  =-\frac{1}{2}\int A_{\mu a}(x)\,D(x,y)^{\mu}{}_{a}|^{\nu}{}_{b}\,A^{\tau}%
{}_{\nu b}(y)\,d^{4}x{}d^{4}y,
\end{align}
where the indices $a$ and $b$ take values in the adjoint representation of the
Lie group of the Yang-Mills theory. The matrix $D$ is given by%
\begin{align*}
D(x,y)^{\mu}{}_{a}|^{\nu}{}_{b}  &  \equiv\delta_{ab}\left(  g^{\mu\nu
}\partial^{2}-\left(  1-\frac{1}{\xi}\right)  \partial^{\mu}\partial^{\nu
}\right)  \delta^{4}(x-y)\\
&  =\frac{1}{(2\pi)^{4}}\int\left[  \delta_{ab}\left(  g^{\mu\nu}%
(p^{2}-i\epsilon)-\left(  1-\frac{1}{\xi}\right)  p^{\mu}p^{\nu}\right)
\right]  e^{ip\cdot(x-y)}{}d^{4}p.
\end{align*}
Thus the difference between the connection propagator and the Yang-Mills field
propagator is just in the coefficients $\delta^{\sigma}{}_{\tau}%
\,\delta^{\varsigma}{}_{\rho}$ and $\delta_{ab}.$ The type of diagrams one can
construct in either theory are the same, and the only differences between them
are the coupling constants and the statistical factor (whose value is due to
the coefficients $\delta^{\sigma}{}_{\tau}\,\delta^{\varsigma}{}_{\rho}$ and
$\delta_{ab}$). That is, both the topological structure as well as the
algebraic structure of the propagators of each diagram in each theory are the
same: diagrams of each theory only differ in their coefficients. For each
diagram in the gravity theory there is a similar diagram in a Yang-Mills
theory, with the same topological structure and degree of divergence. In
particular, the counterterm diagrams in the Yang-Mills theories have
counterparts in the gravity theory. We conclude that the gravity theory we are
discussing should contain a counterterm for every divergent term (using only a
finite set of primitive diagrams) since Yang-Mills theories do. They should be renormalizable.

We have to go back and explain why we ignored the second term of (\ref{lag4}):%
\[
\frac{2}{\kappa^{2}}\left(  g^{\sigma\nu}\Gamma^{\rho}{}_{[\nu\sigma,\rho
]}+g^{\sigma\nu}\Gamma^{\rho}{}_{[\rho\eta}\Gamma^{\eta}{}_{\sigma]\nu
}\right)  .
\]
This term is made up of two possible vertices. Quantity $2\kappa^{-2}%
g^{\sigma\nu}\Gamma^{\rho}{}_{[\nu\sigma,\rho]}$ is a tadpole diagram. Let us
assume that the theory of gravity that we have constructed with the first term
of (\ref{lag4}) is renormalizable. Now, using this one-point vertex it would
be possible to take a connection line (that is, a connection propagator) and
cut it, so it becomes two tadpoles. There is an old (fairly evident) theorem
in renormalization theory that says that, given a renormalized diagram, the
diagram obtained by impeding the flow of momentum through one of the lines is
also renormalized. So new diagram we have obtained using the tadpoles is
renormalized if the first one was. Quantity $2\kappa^{-2}g^{\sigma\nu}%
\Gamma^{\rho}{}_{[\rho\eta}\Gamma^{\eta}{}_{\sigma]\nu}$ is a two-point vertex
and it gives a correction on a line, in a way similar to a mass correction. If
one takes a renormalizable diagram and applies this vertex to one of the
lines, it cuts this line and makes into two, so that the connection propagator
is squared, but no extra loop integration is added. Since the diagrams of this
theory only diverge logarithmically and linearly, adding an extra propagator
that goes as $q^{-2},$ where $q$ is the transferred momentum, will reduce the
divergence by two powers, and make the resulting integration non-divergent. So
the new diagram does not diverge.

\section{Comments.}

We introduced a version of gravity characterized by a term quadratic in the
Riemann tensor in addition to the usual Hilbert-Einstein term, and an affine
connection that is taken not to be Levi-Civita, but a field independent of the
metric. See (\ref{lag2}). It is then shown that for low densities of matter
and energy the model is equivalent to the GTR.

We then studied the renormalization of this theory. It is pointed out that the
resulting action does not contain partials of the metric, a fact that implies
that the metric is a classical field and that there are no propagator lines
due to the metric in the Feynman diagrams of the theory. Thus the only quantum
field in this theory is the connection $\Gamma^{\rho}{}_{\mu\sigma}$; the
metric $g_{\mu\nu}$ is just a background classical field. It is also pointed
out that the vertices of more than two points of this theory appear only on
the term quadratic in the Riemann tensor of the Lagrangian, and not in the
linear one. (Remember the connection is the only quantum field present.)

We conclude that this theory of gravity and Yang-Mills theories have the same
topological and algebraic structure for their diagrams, and that their
respective diagrams differ only on numerical factors due to the different
constants at the couplings (their respective Kronecker deltas and coupling
constants). Since Yang-Mills theories are renormalizable, this theory of
gravity theory should be renormalizable.

\end{document}